\documentclass[a4paper,twocolumn,aps,preprintnumbers,amsmath,amssymb]{revtex4}
\usepackage{color}
\definecolor{Violet}{rgb}{0,0,0.6}

\usepackage[linkcolor=Violet, colorlinks=true, citecolor=Violet, bookmarks=false]{hyperref}

\usepackage{graphicx,rotating}
\usepackage{amsmath,amssymb}
\newcommand{\unit}[1]{\ensuremath{\,\mathrm{#1}}}

\newcommand{\ns}{\unit{ns}\hspace{1mm}}
\newcommand{\us}{\unit{\mu s}\hspace{1mm}}

\newcommand{\MHz}{\unit{MHz}\hspace{1mm}}

\begin{document}

\title{Decoherence and Interferometric Sensitivity of BosonSampling in Superconducting Resonator Networks}
\author{Samuel Goldstein$^1$, Simcha Korenblit$^1$, Ydan Bendor$^1$, Hao You$^2$, Michael R. Geller$^2$ and Nadav Katz$^1$}
\affiliation{$^1$Racah Institute of Physics, The Hebrew University, Jerusalem 91904, Israel}

\affiliation{$^2$Department of Physics and Astronomy, University of Georgia, Athens, Georgia 30602, USA}
\date{\today}
\begin{abstract}
 Multiple bosons undergoing coherent evolution in a coupled network of sites constitute a so-called quantum walk system. The simplest example of such a two-particle interference is the celebrated Hong-Ou-Mandel interference. When scaling to larger boson numbers, simulating the exact distribution of bosons has been shown, under reasonable assumptions, to be exponentially hard. We analyze the feasibility and expected performance of a globally connected superconducting resonator based quantum walk system, using the known characteristics of state-of-the-art components. We simulate the sensitivity of such a system to decay processes and to perturbations and compare with coherent input states.
\end{abstract}
\maketitle

Superconducting Josephson devices are a remarkable quantum information processing platform, with single and two qubit coherences close to and even surpassing fault tolerant thresholds \cite{barends2014superconducting, ofek2016extending}. In harmonic superconducting resonators, various non-classical states have been formed on-demand and complex entangled states between such resonators have also been demonstrated, including NOON states of high order \cite{PhysRevA.85.022335, lang2013correlations, wang2011deterministic}. It is very important to benchmark the quality of entanglement achieved and to steadily expand the size of the systems under study. As such entanglement grows larger, it becomes a resource for potential sensing applications \cite{PhysRevLett.56.1515, Jones29052009} and, more fundamentally, challenges various models of spontaneous collapse \cite{PhysRevLett.44.1323, PhysRevA.45.5193} or correlated error \cite{PreskillProCon}.

A relatively simple (experimentally) platform for such explorations is the recently reformulated problem of quantum walks and associated BosonSampling \cite{Aaronson}. Indistinguishable bosons are placed in a coupled array of resonators and allowed to interfere via the non-interacting coherent quantum walk of the particles among the resonators.
Assuming a closed system (without gauge fields), the boson network then evolves in time with the Hamiltonian
\begin{equation}
	H=\hbar\sum_{i,j}J_{ij}\hat a_i ^\dagger \hat a_j
\end{equation}
where the summation is over all nodes in the graph, $J_{ij}$ is the coupling strength between node $i$ and $j$, and $\hat a_i$ is the ladder operator for resonator $i$. Terms for which $i=j$ can be included to account for varying resonator oscillation energies\cite{PhysRevLett.114.243601}.

The resulting distribution of occupation probabilities in the different resonators is (thought to be) both exponentially hard to compute and verify classically \cite{Aaronson}. This means the problem belongs to the complexity class \#P, which is considered larger and more difficult than the notorious NP class. Note, however, that BosonSampling is expected to be non-universal in the computational sense. Universality with a multi-particle quantum walk hardware was only recently established in the presence of interactions between the bosons \cite{childs2009universal}. 

Although initially this form of quantum simulation was thought to have no practical applications, it has recently been shown to be capable of simulating the vibronic spectra of molecules, if a non-trivial initial state can be created \cite{huh2015boson}, in contrast, however, with the simple single photon inputs of BosonSampling.

Experimental implementations of such multi-boson interference experiments have been carried out mostly in optical qubit experiments by several groups \cite{tillmann2013experimental, spring2013boson, broome2013photonic, wang2015supremacy}. The main limiting factor in these experiments is the exponential overhead in generating the input few-particle state, due to the lack of a deterministic generation of photons. Low collection and detection efficiencies also limit the rate and scaling. Phonon evolution in ion trap arrays have also been implemented for quantum walks, but the geometry and coupling in this system limit the scaling of what is achieved to only a few particles. A theoretical proposal for gated implementation of BosonSampling for microwave photons in a linear array of superconducting resonators was also recently presented \cite{peropadre2015microwave}, followed by a proof of its supremacy \cite{latmiral2015supremacy}. 

\begin{figure}%
\includegraphics[width=0.45\textwidth]{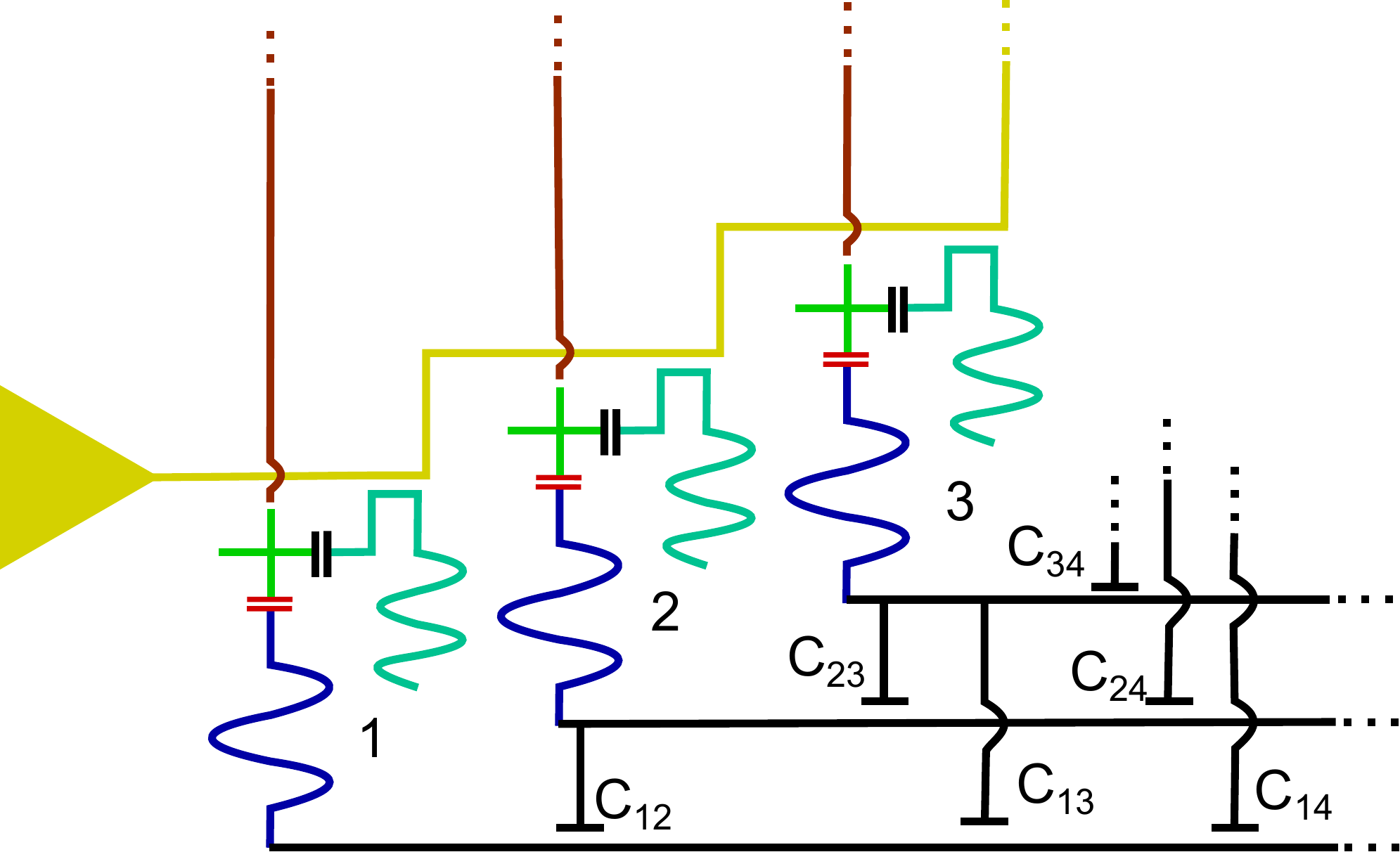}%
\caption{A scheme for implementing BosonSampling. The resonators are loaded from qubits (here represented as the green schematic xmons\cite{barends2013coherent}), and their final state is dispersively measured from readout resonators (magenta) at the desired time when qubits are brought back into resonance with their respective resonators (blue). The resonators are capacitively coupled to each other (black waveguides) with capacitance $C_{ij}$, and this graph is made possible by air-bridges \cite{chen2014airbridge} (arcs on control lines and graph coupling lines) that add a negligible and known capacitive coupling between lines.}
\label{scheme}
\end{figure}

In this paper, we numerically analyze the potential of a superconducting qubit and resonator BosonSampling implementation with current state-of-the-art coherence times and known coupling and readout capabilities. A schematic design and layout of such a system is shown in Fig. \ref{scheme}. We analyze both sensitivity to relaxation and dephasing processes and compare the interferometric sensitivity of the network given either coherent states or single photons as inputs.

In our proposed quantum walk system, single microwave excitations ('bosons') are loaded into interface qubits, and transferred coherently into the resonator network \cite{johnson2010quantum}. The qubits are tuned away from coupling resonance during the quantum walk evolution of the coupled network. The boson occupation numbers in each resonator are subsequently determined by dispersive microwave measurements on the qubits\cite{wallraff2004strong} after swapping them back at a predetermined time.

The advantages of superconducting BosonSampling lie in the on-demand state preparation and high fidelity readout \cite{PhysRevLett.112.190504} at $\sim 99\%$. Another striking strong point of this implementation is the possibility to arbitrarily couple many or even all the resonators to each other, and not just along a one-dimensional array \cite{underwood2012lattice,peropadre2015microwave}.

In order to numerically analyze this proposal we have simulated the state preparation (boson loading) fidelity and the subsequent evolution fidelity. The Hilbert space of $N$ bosons evolving in $M$ resonators (including all decay channels) is of dimension  $D=\binom {N+M}{M}$ , so $D\approx 3\times10^7$ for $N=10$ and $M=20$, and as the system evolves in time full matrices must be used with $\approx 10^{15}$ entries per time step! Some features can be approximated by contour integrals methods \cite{opanchum2016quantumsoftware}, but not all. Thus, a sufficiently high fidelity BosonSampling machine simulation will quickly grow beyond the simulation capabilities of classical computers. Furthermore, as an open physical system it will also be exposed to errors (in particular decoherence effects and perturbations to network couplings). It is the main purpose of this paper to quantify the errors accumulated, subsequently defining the limits of performance for the proposed BosonSampling device. 

In order to reach this goal, we must first determine the necessary simulation duration in order to produce non-trivial results. This is the time period needed for a single boson evolving in the array to establish significant density matrix components throughout ($\sim$25 \ns in this case), and we term this value the "Richness Time".  Hence, a sufficiently complex boson pattern is expected since all the particles have a significant chance of meeting throughout the array, with subsequent multi-particle interference.

We define the richness time in terms of the variance of the expected occupations of all possible states accessible to the system by the requirement
\begin{equation}
	Var(P_i(T_{rich}))=1/M^2
\end{equation}
where $P_i(t)$ is the occupation probability of each state as a function of time $t$.

\begin{figure}%
\includegraphics[trim=3cm 8cm 3cm 8cm, clip=true, width=0.45\textwidth]{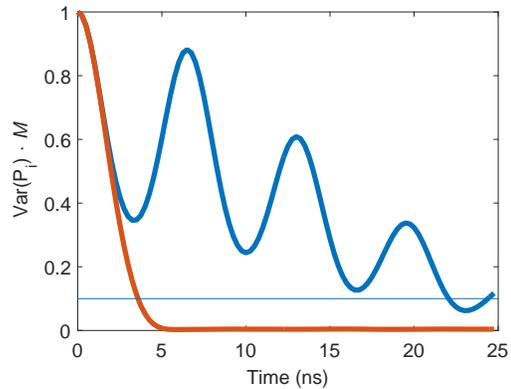}%
\caption{Normalized variance of the occupation probabilities for a single boson in the network $Var(\left\langle P_i(t)\right\rangle) \cdot M$ for $M=10$ (red line for $M=250$) randomly and globally connected resonators. The horizontal line ($\frac{1}{M}$ for $M=10$)) intersects the blue curve at $T_{rich}$. Note that the normalized variance can rise again over $1/M$ as there is no decay in this simulation. For 250 resonators the richness time is reduced to $\sim 5$ \ns.}%
\label{fig:richness200}%
\end{figure}

In Fig. \ref{fig:richness200} we plot the evolution of the normalized variance for a single boson coherently diffusing in the network for two different network sizes ($M=10$ and $M=250$). We have found $T_{rich}$ to decrease slowly with the number of resonators. This can be intuitively understood - as the number of resonators coupled to the initially populated resonator grows, we expect the initial mode to be depleted more quickly. 

Having established the evolution time necessary for sufficient mixing of the multi-particle walk, we now turn to the main results of this work. In all numerical experiments, we calculate only a moderate exponential increase in the distribution distance $\Delta(\rho(t),\sigma(t))$ over time, where $\rho(t)$ and $\sigma(t)$ are the density matrices after coherent and decohered time-evolutions respectively. The distribution distance is defined $\Delta(\rho,\sigma)=\left\|D_{\rho}-D_{\sigma}\right\|_1 $, i.e. the 1-norm distance between the vectors $D_{\rho}$ and $D_{\sigma}$, which represent probability distributions in $\rho$ and $\sigma$ (the matrix diagonals). We find this quantity to fit the phenomenological formula: 

\begin{equation}
\Delta(\rho(t),\sigma(t))= 2(1-e^{-\frac{Nt}{3}(\frac{5}{2}\frac{1}{T_1}+\frac{1}{T_\phi})})
\label{eq:distdistance}
\end{equation}
where $T_1$ and $T_\phi$ are the energy relaxation and pure dephasing times of the resonators.  High quality, planar, state-of-the-art superconducting resonators and qubits have demonstrated energy decay times $T_1$ of $\approx 50$ \us and $T_\phi$, the dephasing time, is here conservatively assumed to be $\approx 50 $\us \cite{Pappas2013}. 

An alternative quantitative estimator of the effects of decoherence is the trace distance, $\mathcal{D}(\rho,\sigma)$. Experimentally, the trace distance is not as easily measured as $\Delta(\rho,\sigma)$, but allows some deeper insight to the evolution of the quantum phases than $\Delta(\rho,\sigma)$. This and other features of the trace distance are discussed in the supplementary material \cite{supmat01}.

We observe in our simulations that the number of resonators does not appear directly to affect the decay, and only enters implicitly through the requirement of achieving the richness time of fully propagating individual bosons throughout the array.
\begin{figure}
\includegraphics[trim=4cm 8.5cm 4cm 8cm, clip=true,width=0.5\textwidth]{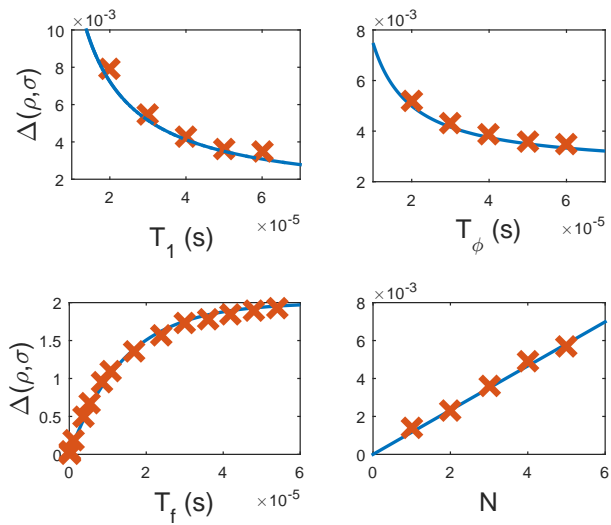}
\caption{Distribution distance, $\Delta(\rho,\sigma)$ for a decohering (with relaxation $T_1=50 \mu s$ and dephasing $T_\phi=50 \mu s$ processes) quantum walk of $N=3$ bosons in a 10 site resonator network, simulated up to $T_f=25 ns$. in each panel plot we simulate the outcome of varying one of the above default values and compare to the phenomenological equation \ref{eq:distdistance}. top left: varying $T_1$, top right: varying $T_\phi$, bottom left: varying $T_f$, bottom right: varying $N$. }
\label{fig:distance}%
\end{figure}

The reliability of this empirical approximation can be seen in Fig. \ref{fig:distance}, as we scan the different parameters of our simulation and plot Eq. \ref{eq:distdistance}. For all panels (unless the parameter is explicitly scanned in the x-axis), we set $T_1=50 \us$, $T_\phi=50 \us$, three excitations, ten resonators, and random global couplings with a uniform distribution between 20 and 40 \MHz. The same randomly generated coupling graph was used in all simulations (iterations performed using other randomly generated graphs with similar energies did not reveal any remarkable difference). The exponential growth of the required Hilbert space required extensive runs (over a few hours per run) on a powerful desktop computer to complete simulations for $N>4$.

Extrapolating to larger resonator arrays and boson numbers, we observe that the mild effects of dephasing allow realistically extending to $N$ of order 20 with currently available coherence times, clearly growing beyond the capabilities of modern classical supercomputing. We note that the fidelity will not be limited by decoherence and decay, rather by loading time for the resonators, initial state fidelity and/or readout fidelity, which are still being developed. 

The mild scaling with $N$ is a somewhat surprising result, as the dephasing Kraus operators cause an $N^2$ faster loss rate of fidelity for a superposition of $N$ bosons and the vacuum state in a single resonator. In this model, however, the initial state diffuses on a very fast time scale between all the modes and the overall state seems to be resistant to dephasing. This is reminiscent of the recent result of Motes et. where BosonSampling was investigated in the context of metrology \cite{motes2015linear}. In addition, the lack of interactions obviously limits the exploited phase space and degree of entanglement in such a system.

\begin{figure}%
\includegraphics[trim=4cm 8cm 4cm 8.5cm, clip=true,width=0.5\textwidth]{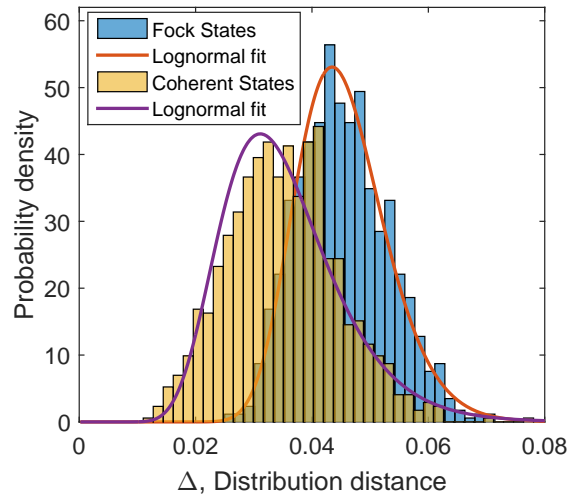}
\caption{Histogram of the distribution distances, $\Delta$, for initial single-photons in resonators and coherent states. The simulation was carried out in a 10 site network, with an ensemble of perturbations, and $N=3$ bosons. The same perturbations were applied for both kinds of initial states. For the coherent state a classical "amplitude simulation" was performed, with $\alpha$=1 in three resonators in the initial state.}%
\label{fig:DistDistHistogram}%
\end{figure}

We test the sensitivity of our system by randomly perturbing the couplings between resonators. An ensemble of perturbation graphs is generated (again with random couplings uniformly chosen between 20 and 40 MHz) and added to the reference graph with a pre-factor of $10^{-3}$. This procedure establishes a slightly altered Hamiltonian by which the system evolves - here without decoherence operators. The subsequent distribution distance between a perturbed and reference evolution after a simulation of $2 T_{rich}$ is calculated. Specifically, we take a random-coupling network of $10$ resonators and generate 1000 perturbed graphs.

The distribution distances yielded by perturbations in the graphs are now computed for both initial Fock states (one photon in each of three different resonators) and initial coherent states (with an average occupation of one photon in three of the resonators) to distinguish classical interferometric sensitivity vs. many-body effects.

Fig. \ref{fig:DistDistHistogram} shows the distribution distances for many realizations of perturbed couplings, and we observe a definite difference between the coherent state inputs and single photons. Both histograms of distribution distances are fit well by lognormal distributions, but the input Fock state is less robust to perturbations than the coherent states. This is promising as it indicates higher interferometric sensitivity for the quantum walk interferometry vs. classical (coherent state) probing. Surprisingly, the result is the opposite for the trace distance metric described above \cite{supmat01}.

\begin{figure}%
	\includegraphics[trim=2cm 8cm 3cm 8.5cm, clip=true, width=0.45\textwidth]{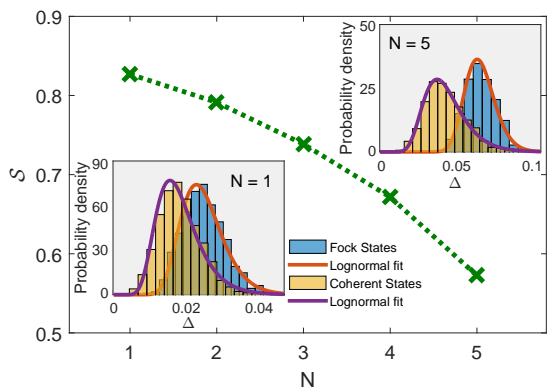}%
	\caption{The distribution overlap, $\mathcal{S}$ as defined in eq. \ref{eq:distoverlap} vs. $N$, the number of bosons, decreases, as $N$ grows. Inserts visualize $\mathcal{S}$ and this effect as the shaded region for $N=1$ and $N=5$.}
	\label{fig:DistributionOverlap}
\end{figure}

The higher sensitivity of the Fock states is enhanced, as the number of bosons increases. To quantify this effect, we define the "distribution overlap":
\begin{equation}
\mathcal{S}=\int dx \sqrt{f_{Fock}f_{coh}}
\label{eq:distoverlap}
\end{equation}
where $f_{Fock}$ and $f_{coh}$ are the probability distributions emerging from iterations with initial Fock states and coherent states respectively, as those in Fig. \ref{fig:DistDistHistogram}. $\mathcal{S}$ equals unity, when the two distributions are identical. Fig. \ref{fig:DistributionOverlap} shows the decrease of $\mathcal{S}$ with growing $N$. 

Another important aspect of the analysis is to address the problem of BosonSampling verification. Obviously the classical computer simulation can serve for verifying the proper distribution and correlations in small boson number implementations. For large boson numbers, the system may be treated classically \cite{seshadreesan2015boson}. However, in the intermediate regime quantum effects dominate, and a classical computer will fail. 

We note also connections to the extensively investigated Anderson localization phenomena, disorder causes a transition from ballistic propagation to localization of particles traveling in a lattice. This has been demonstrated with photons travelling in a two dimensional coupled lattice of waveguides, a system closely related to our proposal. \cite{segev2013anderson, schwartz2007transport}.

For our fully connected graph, symmetry and interference of the propagating excitations creates the exact opposite effect. When disorder is removed from the system and all couplings are made equal, the excitations become confined to their initial resonators since they are detuned from the dressed reservoir of other sites - it is the disorder in interaction strengths that releases them. This effect becomes more pronounced as the number of resonators is increased, starting with simple sloshing between two modes. For the same reasons, when initially placing an equal amount of excitations in each resonator, there is no change in the occupation of the resonators. This is of course true even with disorder.

Our results bode well for a continuous quantum walk implementation of BosonSampling on superconducting Josephson devices with currently achievable lifetimes. From the above analysis, it seems that building a BosonSampling device capable of calculations beyond the abilities of classical computers is within reach.
\begin{acknowledgements}
	This work is supported by the Lady Davis Foundation and the European Research Council project number 335933.
\end{acknowledgements}

\end{document}


\title{Decoherence and Interferometric Sensitivity of BosonSampling in Superconducting Networks - Supplementary materials}
\author{Samuel Goldstein$^1$, Simcha Korenblit$^1$, Ydan Bendor$^1$, Hao You$^2$, Michael R. Geller$^2$ and Nadav Katz$^1$}
\affiliation{$^1$Racah Institute of Physics, Hebrew University of Jerusalem, Jerusalem 91904, Israel}

\affiliation{$^2$Department of Physics and Astronomy, University of Georgia, Athens, Georgia 30602, USA}
\date{\today}
\maketitle

\begin{center}
	\textbf{Trace distance}
\end{center}
\renewcommand{\thefigure}{S\arabic{figure}}
\setcounter{figure}{0}
\renewcommand{\theequation}{S\arabic{equation}}
\setcounter{equation}{0} 

\begin{figure}[h]
	\includegraphics[trim=4cm 8.5cm 5cm 9cm, clip=true, width=0.45\textwidth]{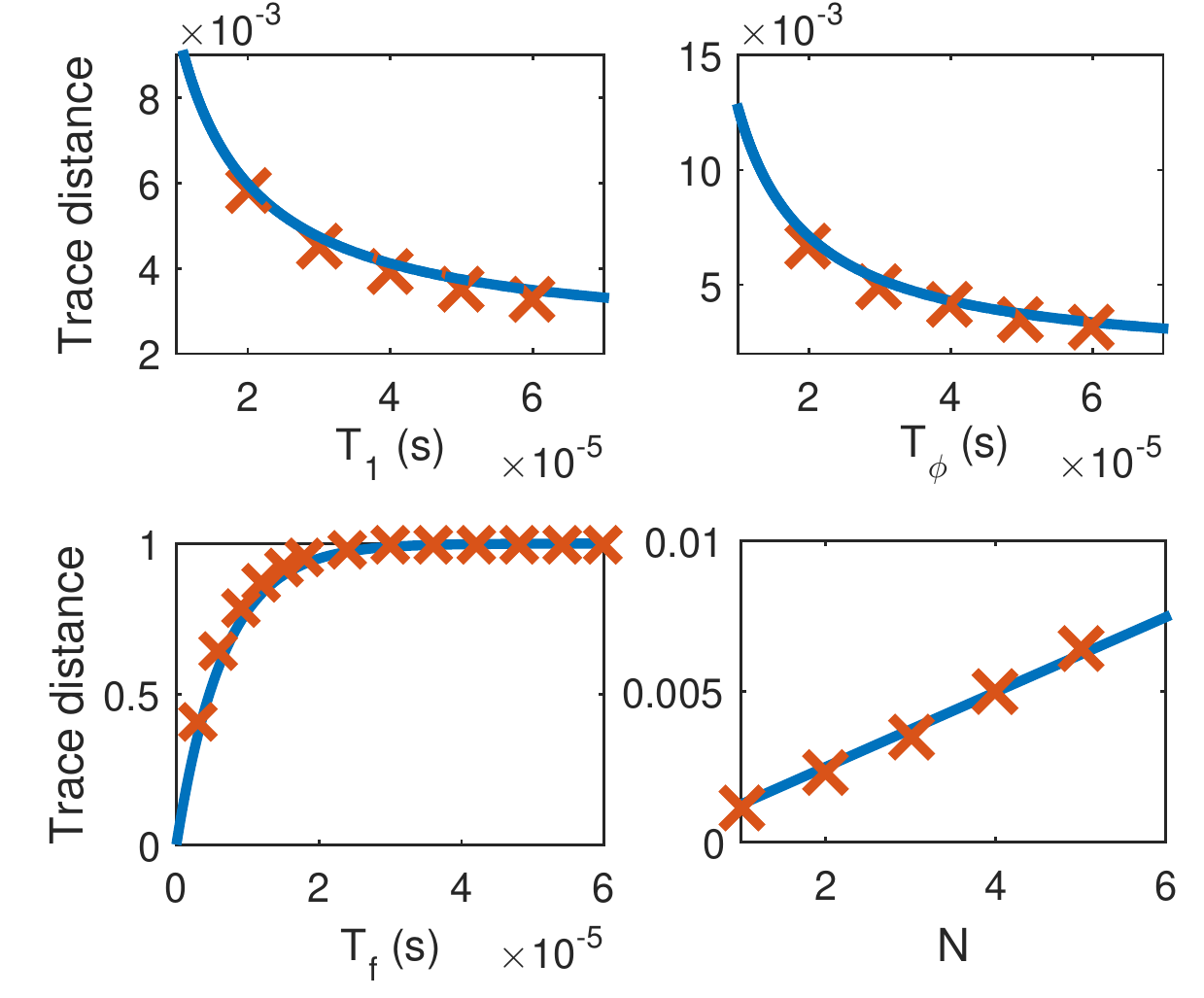}%
	\caption{Trace distance for a decohering ($T_1=50 \mu s$ and $T_\phi=50 \mu s$) quantum walk, parallel to the results presented for the distribution distance in the main article ($N=3$ bosons in a 10 site network, $T_f=25 ns$, with the outcome of varying one of the above values) and comparing to the phenomenological formula \ref{eq:tracedist} (blue curves with no fit parameters). Top left: varying $T_1$, top right: varying $T_\phi$, bottom left: varying $T_f$, bottom right: varying $N$.}
	\label{fig:fig3tracedist}%
\end{figure}

The trace distance $\mathcal{D}(\rho(t),\sigma(t)))$ is a quantitative measure of decoherence, alternative to the distribution distance $\Delta(\rho(t),\sigma(t))$ adopted in the main paper, and it is harder to measure in experiments than $\Delta(\rho,\sigma)$ (but possible with direct Wigner tomography \cite{PhysRevLett.110.100404}). Defined as $\mathcal{D}(\rho,\sigma))= \frac{1}{2}\sum_i |\lambda_i|$, where $\lambda_i$ are the eigenvalues of $(\rho-\sigma)$, we find the trace distance to be well fitted by the model  
\begin{equation}
\mathcal{D}(\rho(t),\sigma(t)))= 1-e^{-Nt(\frac{1}{T_1}+\frac{3}{2}\frac{1}{T_\phi})}
\label{eq:tracedist}
\end{equation}
It is stressed that $\mathcal{D}$ does not depend on the specific couplings between the $M$ resonators (this was verified with different randomly generated graphs). In Fig. \ref{fig:fig3tracedist}  we plot the trace distance under variation of the four parameters, $T_1$, $T_\phi$, $T_f$, and $N$. All of the above were also altered in the main paper for the analysis of the distribution distance, and the results are compared to the phenomenological model of Eq. \ref{eq:tracedist}. 
Just as the case for $\Delta(\rho,\sigma)$, we observe that the errors displayed in Fig. \ref{fig:fig3tracedist} due to decoherence (the two top panels) are in order of less than one percent. However, we observe $\mathcal{D}(\rho,\sigma)$ to be more sensitive to changes in $T_{\phi}$ than to $T_1$, and more sensitive to such changes in $T_{\phi}$ than $\Delta(\rho,\sigma)$ was found to be. On the other hand, $\Delta(\rho,\sigma)$ was more sensitive to changes in $T_1$. Intuitively this conclusion can be understood by the definition of $\mathcal{D}(\rho,\sigma)$, where off-diagonal terms are considered and thus relating more directly to the dephasing effects, i.e. $T_{\phi}$. Since $\Delta(\rho,\sigma)$ measures the probabilities only and not the phases directly, the energy relaxation implemented by $T_1$ naturally has a more significant effect on this parameter than $T_{\phi}$.

\begin{center}
	\textbf{Bound on distribution distance}
\end{center}

\begin{figure} [b]
	\includegraphics[trim=2cm 7cm 3cm 7.5cm, clip=true, width=0.45\textwidth]{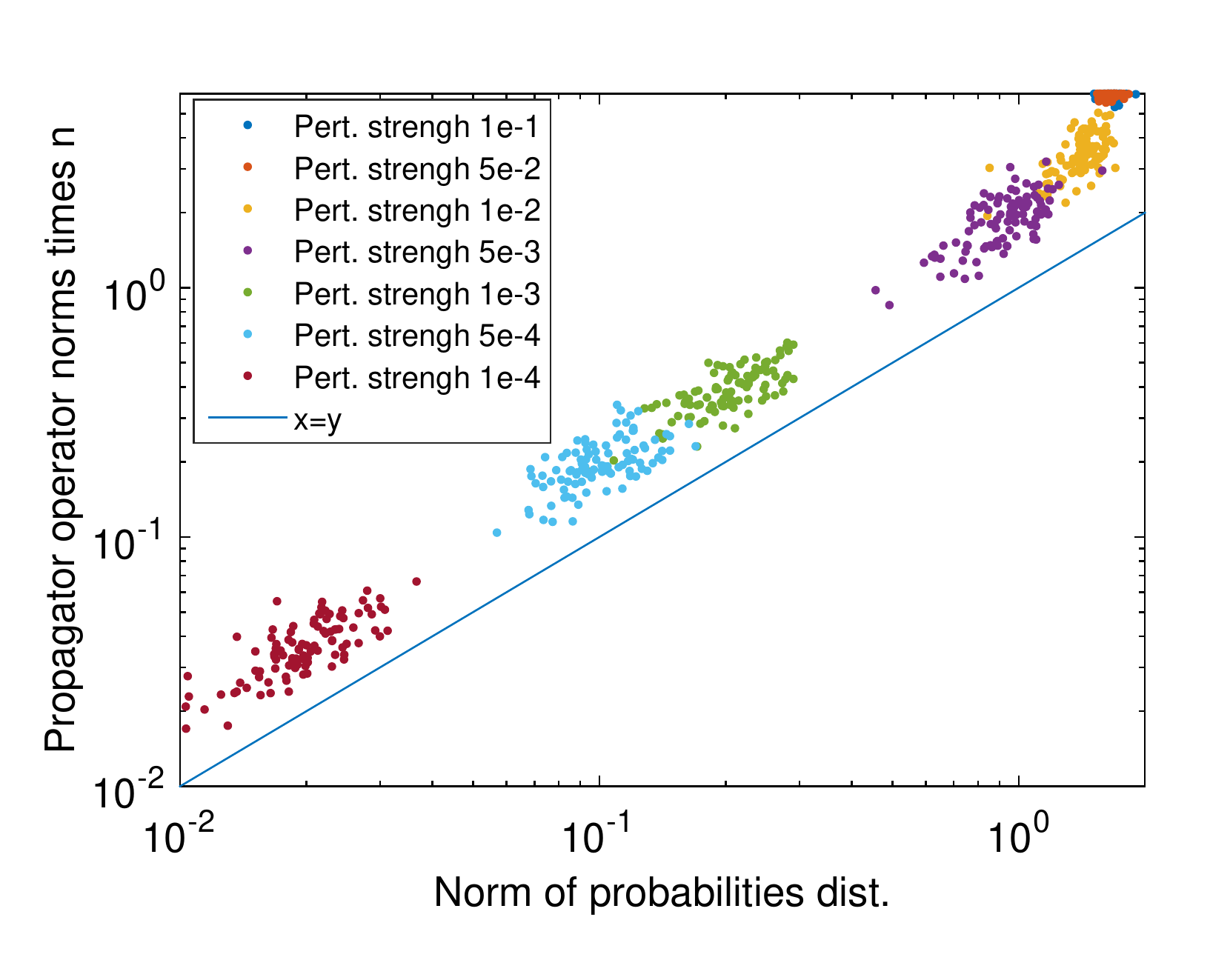}
	\caption{The norm of the distribution error was compared to the operator norm of the propagator perturbation time $N$ (three in all cases) as in eq. \ref{eq:ArkhipovEq} for 200 perturbations at four strengths. In all iterations the latter is larger than the former, leaving all dots above the line of equality.} 
	\label{fig:Arkhipov}%
\end{figure}
In a recent theoretical work by Arkhipov \cite{arkhipov2015bosonsampling}, an upper bound on the error of the boson sampler's output distributions is presented. This bound is shown to depend on the deviation in the unitary time propagator $U=e^{iGt}$ (where $G$ is the coupling graph between the modes in the network), and is mathematically express by \cite{arkhipov2015bosonsampling} as:

\begin{equation}
\|D_{\tilde{U}}-D_{U}||_1 \leq N ||{\tilde{U}}-U||_{op}
\label{eq:ArkhipovEq}
\end{equation}

Here $\|D_{\tilde{U}}-D_{U}||_1$ is the 1-norm distance between the sampled distributions, and the operator distance is defined by \cite{arkhipov2015bosonsampling} as $||{\tilde{U}}-U||_{op}=\max_{i}|\lambda_{i}-1|$, where $\lambda_i$ are eigenvalues of $\tilde{U}U^{-1}$.  

An analytic proof was provided by \cite{arkhipov2015bosonsampling}. In Fig. \ref{fig:Arkhipov} we confirm the statement numerically with our simulation platform. In this section we perturbed the network's coupling graph and consequently the propagator, and compared the norm of the resultant error of the outcome distributions (left-hand-side of Eq. \ref{eq:ArkhipovEq}) to the appropriate operator norm of the propagator perturbations (the right-hand-side of eq. \ref{eq:ArkhipovEq}). This was done for 200 perturbations at four different strength, and for none of these iterations Eq. \ref{eq:ArkhipovEq} was violated. Furthermore, we find the bound to be tight for our system.

\begin{center}
\textbf{Perturbation sensitivity}
\end{center}

\begin{figure}[h]
	\centering
	\includegraphics[trim=5.5cm 10cm 6.5cm 10.5cm, clip=true, width=0.40\textwidth]{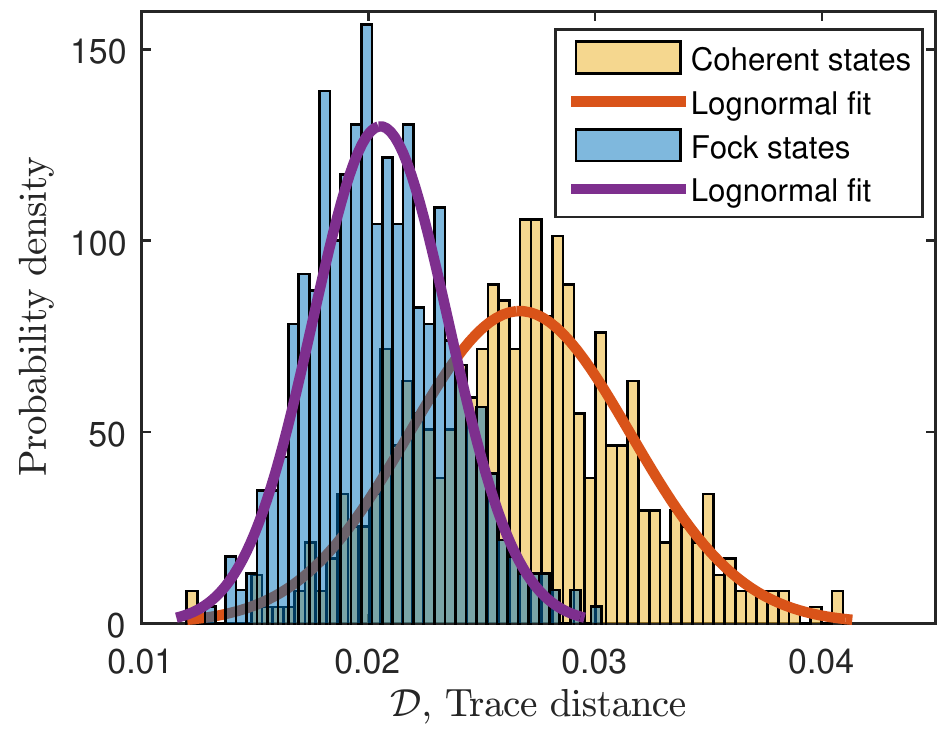}
	\caption{Distribution of the trace distance, $\mathcal{D}$ for initial single-photons in resonators and coherent states. The simulation was carried out in a 10 site network, with an ensemble of perturbations and N=3 bosons. For the coherent state a classical "amplitude simulation" was performed, with $\alpha$=1 in three resonators in the initial state.}
	\label{fig:TraceHist}%
\end{figure}

In our paper, our study of the perturbed coupling graphs proved input Fock states to be more sensitive to perturbations than coherent states. This was concluded by means of the distribution distance, but again the trace distance serves as an alternative parameter for comparison between density matrices.

Here we use the overlap $\mathcal{O}$, which in the case of pure states is related to the trace distance by $\mathcal{D}=\sqrt{1-\mathcal{O}^{2}}$ (since such states have $\rho=|\psi\rangle\langle\psi|$ \cite{PhysRevA.93.012110}). In this case, the overlap is conveniently calculated as the squared dot product between the original and perturbed states, i.e. $\mathcal{O}=|\langle\psi|\phi\rangle|$, and $\mathcal{O}$ therefore provides the trace distances with minimum computational effort.

As in the main paper, we take a particular random-coupling $10$ resonator network and generate an ensemble of slightly perturbed graphs, which depends on the number of sites, bosons and the exact strength of the perturbation. 

When we plot the distribution for 1000 random realizations of perturbed couplings in Fig. \ref{fig:TraceHist} we observe a definite difference between the coherent state inputs and single photons - similar to what was displayed in the main paper's discussion of distribution distances. The distributions for both Fock states and coherent states of observed $\mathcal{D}$'s are fit well by a normal distribution (whose source is the random normal perturbations of the coupling graph), but this time it is the input \textit{coherent} states, which are remarkably less robust to noise (larger and broader distribution) contrary to the former conclusion gained with distribution distances. 

This result seems to indicates less interferometric sensitivity for the quantum walk interferometry vs. classical (coherent state) probing, disagreeing with what we found in the main paper. But the explanation as to this difference lies within the nature of coherent states. These dephase more rapid than the Fock states, and this is exactly the tendency, which the trace distance detects advantageously compared to the distribution distance. The decoherence estimator must therefore be chosen appropriately, when given processes are to be distinguished.